\documentclass[twocolumn, conference]{IEEEtran}
\IEEEoverridecommandlockouts
\usepackage{cite}
\usepackage{amsmath,amssymb,amsfonts}
\usepackage{graphicx}
\usepackage{textcomp}
\usepackage{xcolor}
\usepackage{multirow}
\usepackage{subfigure}
\usepackage{algorithm}
\usepackage{algpseudocode}
\usepackage{url}
\usepackage{balance}

\title{DeScoD-ECG: Deep Score-Based Diffusion Model for ECG Baseline Wander and Noise Removal}

\begin{document}

\author{
    \IEEEauthorblockN{Huayu Li, Gregory Ditzler, Janet Roveda and Ao Li}
    \thanks{H. Li is with the Department of Electrical \& Computer Engineering at the University of Arizona, Tucson, AZ 85719 USA. E-mail: hl459@arizona.edu}
        \thanks{G. Ditzler is with the Department of Electrical \& Computer Engineering at Rowan University, Glassboro, NJ 08028 USA. E-mail: ditzler@rowan.edu}
        \thanks{J. Roveda is with the Department of Electrical \& Computer Engineering, Department of Biomedical Engineering, and Bio5 Institute at the University of Arizona, Tucson, AZ 85719 USA. E-mail: meilingw@arizona.edu}
        \thanks{A. Li is with the Department of Electrical \& Computer Engineering and Bio5 Institute at the University of Arizona, Tucson, AZ 85719 USA. E-mail: aoli1@arizona.edu}
}


\maketitle

\begin{abstract}

\textit{Objective:} Electrocardiogram (ECG) signals commonly suffer noise interference, such as baseline wander. High-quality and high-fidelity reconstruction of the ECG signals is of great significance to diagnosing cardiovascular diseases. Therefore, this paper proposes a novel ECG baseline wander and noise removal technology. \textit{Methods:}  We extended the diffusion model in a conditional manner that was specific to the ECG signals, namely the Deep Score-Based Diffusion model for Electrocardiogram baseline wander and noise removal (DeScoD-ECG). Moreover, we deployed a multi-shots averaging strategy that improved signal reconstructions. We conducted the experiments on the QT Database and the MIT-BIH Noise Stress Test Database to verify the feasibility of the proposed method. Baseline methods are adopted for comparison, including traditional digital filter-based and deep learning-based methods. \textbf{Results:} The quantities evaluation results show that the proposed method obtained outstanding performance on four distance-based similarity metrics with at least 20\% overall improvement compared with the best baseline method. \textit{Conclusion:} This paper demonstrates the state-of-the-art performance of the DeScoD-ECG for ECG baseline wander and noise removal, which has better approximations of the true data distribution and higher stability under extreme noise corruptions. \textit{Significance:} This study is one of the first to extend the conditional diffusion-based generative model for ECG noise removal, and the DeScoD-ECG has the potential to be widely used in biomedical applications.
\end{abstract}

\begin{IEEEkeywords}
ECG signal processing, Baseline wander, diffusion models
\end{IEEEkeywords}

\section{Introduction}

\IEEEPARstart{C}{ardiovascular} 
diseases are the leading cause of sudden cardiac death in the world~\cite{world2014global}. Thus, specialized medical services such as diagnostic tools for the study and treatment are in great demand. The electrocardiogram (ECG) is the most straightforward, non-invasive technique and monitoring approach for diagnosing heart diseases. An ECG records the electrical signal from the heart to check for different heart conditions. ECG signals are typically collected during exercise stress tests or in an ambulatory way to increase diagnostic sensitivity. However, the ECG signals acquired in these conditions are strongly influenced by different types of noise, such as baseline wander. The noisy ECG signals are not conducive to both manual and automatic signal analysis, which significantly reduces the usability of the collected signals and increases the cost of clinical experiments and diagnosis.

High-quality and high-fidelity noise removal in the ECG signals is of great significance to the auxiliary diagnosis of heart diseases via ECG. Several solutions have been proposed to restore the ECG signals through a signal processing perspective. Classical digital filter-based methods use finite impulse response (FIR) and infinite impulse response (IIR) filters to remove the noise presented in certain frequencies~\cite{kumar2015removal}. Wavelet transforms decompose the ECG signals into different frequency sub-bands and take advantage of the energy of the signals in different scales to isolate baseline wander from ECG signals. In~\cite{barati2006baseline}, independent component analysis (ICA) is used for removing baseline wander from ECG under the assumptions of different statics between the noise and original ECG signals. Unfortunately, the traditional methods typically can only succeed in situations where the ECG is not corrupted severely and fail when the noise has high amplitude.

Recently, data-driven methods which take advantage of deep neural networks have been used to remove the baseline wander. Work has shown that the neural networks trained on clean and noisy ECG records surpass the performance of the traditional methods by a significant difference. In~\cite{antczak2018deep}, a novel approach to denoise ECG signals with deep recurrent denoising neural networks (DRNN) is proposed. By modeling the ECG time series with an LSTM module~\cite{hochreiter1997long}, DRNN achieved outstanding results compared to the traditional digital filters. Further, the fully convolutional denoising autoencoder (FCN-DAE)~\cite{chiang2019noise} was proposed to denoise the ECG signals via denoising autoencoder with hourglass architecture. The denoising autoencoders~\cite{vincent2008extracting} first learn architectures to extract the most relevant features automatically and then produce the reconstructions in the presence of noise at the input. Recently, a new deep neural network architecture named DeepFilter~\cite{romero2021deepfilter} was presented and achieved state-of-the-art performance that beat the traditional digital filters and other deep learning-based methods. It is inspired by the designation of the Inception architecture~\cite{szegedy2015going} and further proposed a Multi-Kernel Linear And Non-Linear (MKLANL) filter module for denoising the ECG signals with multi-scale features. Also, the generative model has been applied to ECG denoising. An ECG signal denoising method Using Conditional Generative Adversarial Net (CGAN) is proposed in \cite{wang2022ecg}. It trained a neural network with skip connection (well known as UNet~\cite{ronneberger2015u} architecture) under an adversarial framework for better approximation of the true data distribution.


Unfortunately, the existing deep learning-based methods perform poorly with extremely high noise corruption, which limits the applicability of these methods for physical exercise stress tests. Thus, there is a significant need in the research community to develop more accurate ECG reconstruction techniques and find more precise approximations of the conditional distribution of the clean ECG records, given the noisy observations. 
In this work, we propose DeScoD-ECG, a novel probabilistic baseline wander removal method that learns the conditional distribution with conditional score-based diffusion models~\cite{ho2020denoising,song2020denoising}. We extend the score-based diffusion models into a novel conditional generative model specific for processing ECG signals. Instead of directly denoising/filtering the noisy observations, the DeScoD-ECG iteratively generates signal reconstructions from random Gaussian distributions conditioned on the noisy observations. The experimental results show that the DeScoD-ECG achieves state-of-the-art performance and beats the baseline methods with a large gap. Also, compared with the baseline methods, the DeScoD-ECG is more stable and consistent with different noise levels. As one of the first evaluations of the conditional score-based diffusion model for ECG denoising, our main contributions are: (1) a conditional score-based diffusion model for high-quality and high-fidelity baseline wander removal is proposed. To the best of our knowledge, this is the first study to deploy the score-based diffusion model for ECG signal restoration; (2) state-of-the-art performance was achieved in baseline wander removal on ECG records from the QT Database with baseline wander noise from MIT-BIH Noise Stress Test Database; (3) a self-ensemble strategy was adopted to further improve the reconstruction quality and the algorithm stability under extreme noise conditions.

\section{Related Works}
In this section, we introduce the basic concept of electrocardiogram and the cause of baseline wander. We discussed the significance of baseline wander removal from a clinical point of view. Further, we introduce the Score-based generative models and their advantages compared with other generative models.

\subsection{Electrocardiogram and Baseline wander}
An ECG signal records the electrical signals in the heart. It is a common and painless test used to detect heart problems and quickly monitor heart health \cite{zimetbaum2003use}. Electrodes placed at certain spots on the chest and legs are used to detect the potential difference in the human body. An ECG sensor records these impulses to show how fast the heart is beating, the rhythm of the heartbeats (steady or irregular), and the strength and timing of the electrical impulses as they move through the different parts of the heart. Changes in an ECG can be a sign of cardiovascular diseases. An exercise stress test helps determine how well the cardiac responds when working its hardest. It typically involves walking on a treadmill or pedaling on a stationary bike while hooked up to an ECG to monitor the heart activity~\cite{goraya2000prognostic, gosselink2004exercise}. Baseline wander usually appears in the ECG signals during exercise stress tests which could be caused by factors such as respiration, variations in electrode impedance, and excessive body movements. Baseline wander significantly affects the accuracy of analysis and diagnosis from the ECG. Thus, high-quality and high-fidelity removal of baseline wander can improve the efficiency of clinical diagnosis and enable more flexible exercise stress tests so patients can move more freely.

\subsection{Score-based Diffusion model}
Score-based diffusion models are a class of deep generative models that generate samples by gradually converting noise into a plausible data sample through denoising. Score-based diffusion models have recently been used for different tasks such as image generation~\cite{ho2020denoising, song2020score} and text-to-audio synthesis~\cite{kong2020diffwave, chen2020wavegrad} with state-of-the-art sample quality that outperform counterparts including generative adversarial nets (GANs)~\cite{goodfellow2014generative} and variational autoencoders (VAEs)~\cite{kingma2013auto}. Score-based diffusion models address the drawbacks of GANs and VAEs. More specifically, GANs are known for potentially unstable training and less diversity in generation due to their adversarial training nature. VAEs rely on a surrogate loss and have relatively poor generation quality. Unfortunately, tractability and flexibility are two conflicting objectives in generative modeling. Tractable models can be analytically evaluated and cheaply fit data but are unable to describe the structure in rich datasets aptly.
In contrast, flexible models can fit arbitrary structures in data, but evaluating and training these models is more computationally expensive. Diffusion models mitigate the existing drawbacks, which are analytically tractable and flexible. Furthermore, the diffusion models do not need the adversarial training of GAN while having comparable or even better generation qualities than GAN. Therefore, in this work, we introduce the score-based diffusion model to the ECG baseline wander and noise removal tasks and demonstrate the efficiency of this type of generative model.
\section{Methods}

\begin{figure*}
    \centering
    \includegraphics[width=0.9\textwidth]{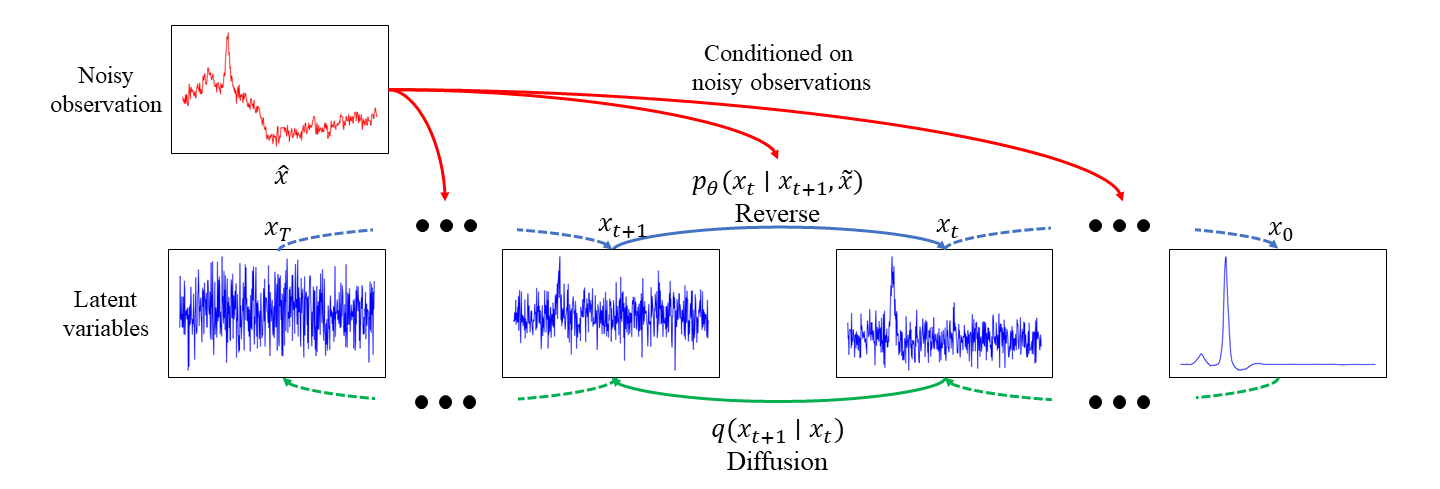}
    \caption{The \emph{diffusion} ($q$) and \emph{reverse} ($p_\theta$) procedure of DeScoD-ECG. Intuitively, the \emph{diffusion} process gradually corrupts the ground truth ECGs by adding noise, while the \emph{reverse} process aims to learn a denoising function to remove this noise. It is noticeable that \emph{reverse} process is conditioned by the corresponding observed ECGs. By training the denoising function (or the model $\theta$) in this way, we can generate the clean ECGs from random Gaussian noise conditioned on the noisy observations.}
    \label{fig:procedure}
\end{figure*}

\begin{algorithm}[t]
\caption{Training.}\label{alg:train}
\textbf{Input}: distribution of clean and noisy ECG pairs $q(x_o, \Tilde{x})$, the number of training iterations $N$, maximal diffusion steps $T$, a predefined noise schedule $S=\left \{1,\sqrt{\bar{\alpha}_0},\dots,\sqrt{\bar{\alpha}_T} \right \}$, 

\textbf{Output}: Trained denoising function $\boldsymbol{\epsilon}_{\theta}$

\begin{algorithmic}[1]
\State \textbf{Initiate}: $i=0$
\While{$i \leq N$}
\State $t\sim\text{Uniform}\left ( \left \{ 1,\dots,T \right \} \right )$
\State $x_o,\Tilde{x} \sim q(x_o, \Tilde{x})$
\State $\bar{\alpha}\sim\text{Uniform}\left ( S_{t-1},S_t \right )$
\State $\boldsymbol{\epsilon}\sim \mathcal{N}(\mathbf{0}, \mathbf{I})$
\State Take gradient descent step on \[\nabla_\theta\left\|\boldsymbol{\epsilon}-\boldsymbol{\epsilon}_{\theta}\left(\sqrt{\bar{\alpha}} x_{0}+\sqrt{1-\bar{\alpha}} \boldsymbol{\epsilon}, \tilde{x}, \bar{\alpha}\right)\right\|^{2}\]
\State $i=i+1$
\EndWhile
\end{algorithmic}
\end{algorithm}

\begin{algorithm}[t]
\caption{Inference.}\label{alg:inference}
\textbf{Input}: Trained denoising function $\boldsymbol{\epsilon}_{\theta}$, a noisy observation $\tilde{x}$, maximal diffusion steps $T$, $M$-shots reconstruction

\textbf{Output}: Reconstructed ECG $x_0$
\begin{algorithmic}[1]
\State \textbf{Initiate}: $x^{m}_0=0$, $m=M$, 
\While{$m>0$}
\State $t=T$, $x_t\sim \mathcal{N}(\mathbf{0}, \mathbf{I})$
\While{$t > 0$}
\State Let 
\[x_{t-1}= \frac{1}{\alpha_t}\left (x_t-\frac{1-\alpha_t}{\sqrt{1-\bar{\alpha}_{t}}}\boldsymbol{\epsilon}_{\theta}\left ( x_t,\tilde{x},\sqrt{\bar{\alpha}_t} \right )  \right )
\]
\If{$t>1$}
\State $x_{t-1}=x_{t-1}+\sigma_t$
\EndIf
\State $t=t-1$
\EndWhile
\State $x^{m}_0 = x^{m+1}_0+x_0$
\State $m=m-1$
\EndWhile
\end{algorithmic}
\textbf{return} $x^{0}_0/M$
\end{algorithm}

\begin{figure*}
    \centering
    \includegraphics[width=.9\textwidth]{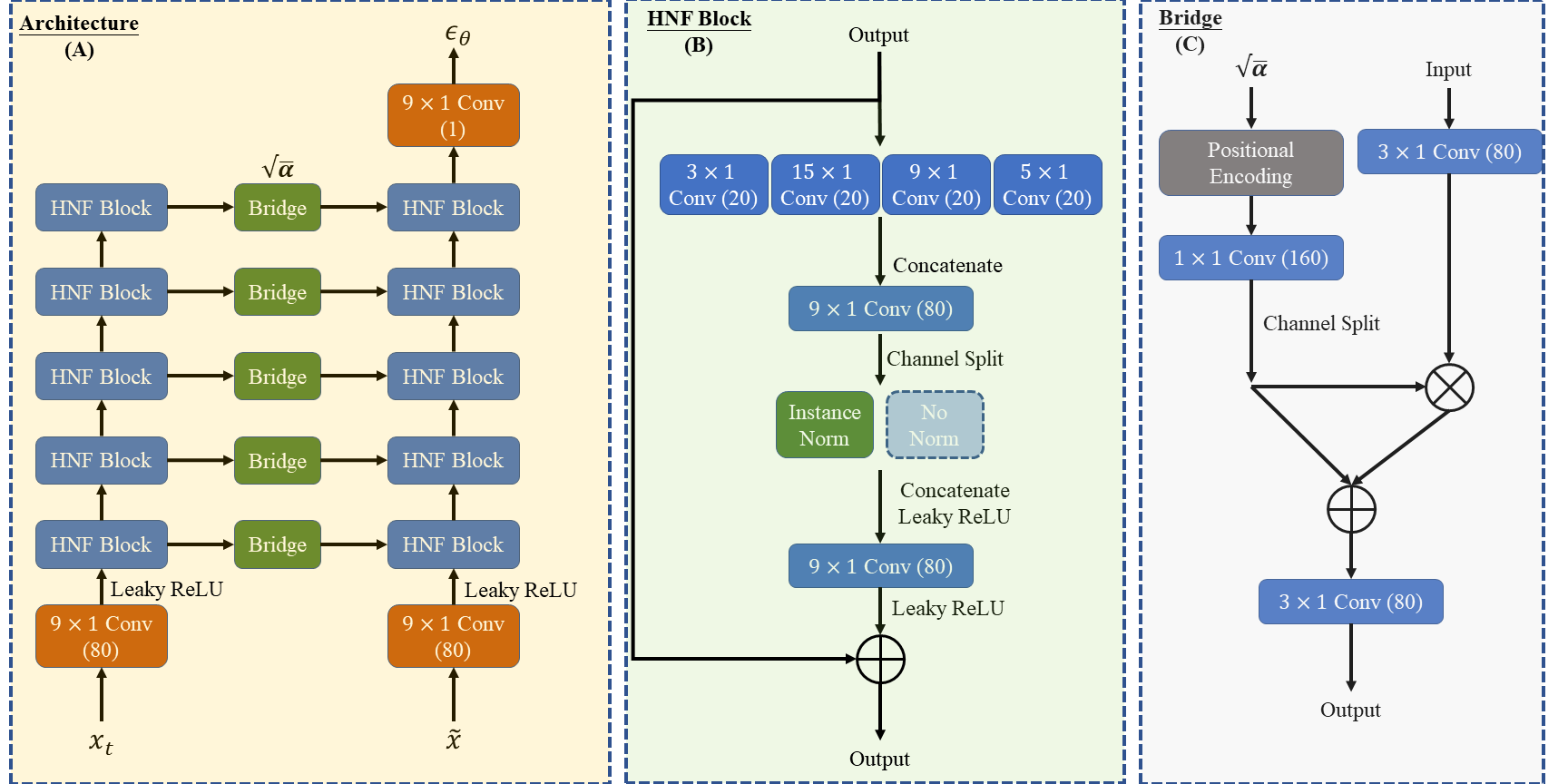}
    \caption{The architecture of the neural network and each block used in it. We demonstrate the network architecture (A) used in DeScoD-ECG. The whole network is made up by two streams equipped with the HNF block (B) and communicated via the Bridge block (C).} 
    \label{fig:architecture}
\end{figure*}

In this section, we first introduce the background of diffusion models and then introduce the proposed DeScoD-ECG for ECG baseline wander and noise removal. We also present the implementation details of the neural network and optimization procedure used in this work. 

\subsection{Score-based Diffusion Model}
Diffusion models are generative models, meaning they are used to generate data similar to the data on which they are trained. Fundamentally, diffusion models work by corrupting training data through the successive addition of Gaussian noise and then learning to recover the data by reversing this \emph{diffusion process}. Thus, a well-trained diffusion model is able to generate data from randomly sampled noise through the learned denoising process.

More specifically, given samples from a data distribution $q(x_0)$, we are interested in learning a model distribution $p_\theta(x_0)$ that approximates $q(x_0)$. Denoising diffusion probabilistic models (DDPMs)~\cite{ho2020denoising, sohl2015deep} are latent variable models that consist of a \emph{forward process} or \emph{diffusion process}, in which the data are diffused to a well-behaved distribution by progressively adding noise, and a \emph{reverse process}, in which noise is transformed back into a sample from the target distribution.

The \emph{diffusion process} gradually adds noise to the data to approximate the posterior $q\left(x_{1:T} \mid x_{0}\right)$, where $x_{1}, \ldots, x_T$ are the latent variables with the same dimensionality as $x_0$. In~\cite{ho2020denoising}, the \emph{diffusion process} is set to a simple parameterization of a fixed Markov Chain with conditional Gaussian translation as each
step:
\begin{align}
    q\left(x_{1: T} \mid x_{0}\right)&:=\prod_{t=1}^{T} q\left(x_{t} \mid x_{t-1}\right),\\
    q\left(x_{t} \mid x_{t-1}\right)&:=\mathcal{N}\left(x_{t} ; \sqrt{1-\beta_{t}} x_{t-1}, \,\,\beta_{t} \mathbf{I}\right),
\end{align}
where $\beta_{1}, \ldots, \beta_{T}$ is a variance schedule (either learned or fixed) which, if well-behaved, ensures that $x_T$ is nearly an isotropic Gaussian for sufficiently large $T$, and $\mathcal{N}(x; \mu, \Sigma)$ is a Gaussian pdf with parameters $\mu$ and $\Sigma$. 
The sampling of $x_t$ at an arbitrary timestep $t$ has the closed-form of $q\left(x_{t} \mid x_{0}\right)=\mathcal{N}\left(x_{t} ; \sqrt{\bar{\alpha}_{t}} x_{0},\left(1-\bar{\alpha}_{t}\right) \mathbf{I}\right)$, where $\alpha_{t}:=1-\beta_{t}$ and $\bar{\alpha}_{t}:=\prod_{s=1}^{t} \alpha_{s}$. Thus, $x_t$ can be expressed directly as: $x_{t}=\sqrt{\alpha_{t}} x_{0}+\left(1-\alpha_{t}\right) \boldsymbol{\epsilon}, \text { where } \boldsymbol{\epsilon} \sim \mathcal{N}(\mathbf{0}, \mathbf{I})$.

The \emph{reverse process} learns to reverse this \emph{diffusion process} to recover $x_0$ from $x_t$. Starting with the pure Gaussian noise sampled from $p(x_T):=\mathcal{N}(x_T, \mathbf{0}, \mathbf{I})$, the \emph{reverse process} is defined by the following Markov chain:
\begin{align}
\label{eq:reverse process}
&p_{\theta}\left(x_{0: T}\right):=p\left(x_{T}\right) \prod_{t=1}^{T} p_{\theta}\left(x_{t-1} \mid x_{t}\right), \quad x_{T} \sim \mathcal{N}(\mathbf{0}, \mathbf{I}) \\
&p_{\theta}\left(x_{t-1} \mid x_{t}\right):=\mathcal{N}\left(x_{t-1} ; \boldsymbol{\mu}_{\theta}\left(x_{t}, t\right), \sigma_{\theta}\left(x_{t}, t\right) \mathbf{I}\right)
\end{align}
where the time-dependent parameters of the Gaussian transitions are learned. In the DDPMs setting, the following specific parameterization of
$p_{\theta}\left(x_{t-1} \mid x_{t}\right)$ is proposed:
\begin{align}
    &\boldsymbol{\mu}_{\theta}\left(x_{t}, t\right)=\frac{1}{\alpha_{t}}\left(x_{t}-\frac{\beta_{t}}{\sqrt{1-\alpha_{t}}} \boldsymbol{\epsilon}_{\theta}\left(x_{t}, t\right)\right),\\
    &\sigma_{\theta}\left(x_{t}, t\right)=\sqrt{\tilde{\beta}_{t}}, \text { where } \tilde{\beta}_{t}= \begin{cases}\frac{1-\alpha_{t-1}}{1-\alpha_{t}} \beta_{t} & t>1 \\ \beta_{1} & t=1\end{cases}
\end{align}
where $\boldsymbol{\epsilon}_{\theta}(\cdot,\cdot)$ is a learnable denoising function which estimates the noise vector $\boldsymbol{\epsilon}$ that was added to a noisy input $x_t$. This parameterization leads to the following alternative loss function:
\begin{align}
    L_{\text {simple }}(\theta):=\mathbb{E}_{t, x_{0}, \epsilon}\left[\left\|\boldsymbol{\epsilon}-\boldsymbol{\epsilon}_{\theta}\left(\sqrt{\bar{\alpha}_{t}} x_{0}+\sqrt{1-\bar{\alpha}_{t}} \boldsymbol{\epsilon}, t\right)\right\|^{2}\right]
\end{align}
This objective can be viewed as a weighted combination of denoising score matching which enables more stable training and better results than the original score matching losses \cite{song2019generative,vincent2011connection}.

\subsection{Conditional Diffusion Model for ECG Baseline Wander and Noise Removal}
\label{sec:conditional_diff}
In the task of removing the baseline wander and noise in ECG signals, we consider recovery of the clean signal $x_0$ given a noisy ECG signal $\tilde{x}$. The noisy ECG signal contains baseline wander and noise of unknown type. Thus, we estimate the true conditional data distribution $q(x_0 \mid \tilde{x})$  via modeling the conditional distribution $p_\theta(x_0\mid\tilde{x})$ with a diffusion model. Then, we extend the Eq.~\eqref{eq:reverse process} to a conditional fashion:
\begin{align}
\label{eq:cond_fashion}
    &p_{\theta}\left(x_{0: T}\mid \tilde{x}\right):=p\left(x_{T}\right) \prod_{t=1}^{T} p_{\theta}\left(x_{t-1} \mid x_{t},\tilde{x}\right) , x_{T} \sim \mathcal{N}(\mathbf{0}, \mathbf{I}), \\
    &p_{\theta}\left(x_{t-1} \mid x_{t},\tilde{x}\right):=\mathcal{N}\left(x_{t-1} ; \boldsymbol{\mu}_{\theta}\left(x_{t}, t\mid\tilde{x}\right), \sigma_{\theta}\left(x_{t}, t\mid\tilde{x}\right) \mathbf{I}\right).
\end{align}
The conditional \emph{reverse process} $p_{\theta}\left(x_{t-1} \mid x_{t},\tilde{x}\right)$ iteratively reconstructs the clean ECG $x_0$ from a Gaussian distribution with the noisy observations as conditions. 
We visualized the diffusion and reverse mechanism of DeScoD-ECG in Figure \ref{fig:procedure}. The conditional \emph{reverse process} $p_{\theta}\left (x_{t-1} \mid x_{t},\tilde{x}\right )$ gradually denoises the Gaussian noise by a small step conditioned on the noisy observations and then produce the clean signals. Algorithm~\ref{alg:train} details the training process of DeScoD-ECG. We first initiate the noise schedule $\beta$s for $T$ steps diffusion in a quadratic form, that the $\left \{ \beta_1,\dots,\beta_n \right \}$ are defined as:
\begin{align}
    &\beta_1 = 0.0001, \beta_T=0.5,\\
    &\beta_t = \left (\frac{T-t}{T-1}\sqrt{\beta_1}+ \frac{t-1}{T-1}\sqrt{\beta_t} \right )^2.
\end{align}
This noise schedule can improve the synthetic quality with fewer diffusion/reverse steps~as prior works have shown \cite{nichol2021improved, song2020denoising}. 
Recall that $\alpha_{t}:=1-\beta_{t}$ and $\bar{\alpha}_{t}:=\prod_{s=1}^{t} \alpha_{s}$, we define a new noise schedule specific for training process that can avoid the denoising function $\boldsymbol{\epsilon}_\theta$ by directly conditioning on the discrete diffusion step index $t$ and will improve the model performance. More specifically, we use the predefined noise schedule $S=\left \{1,\sqrt{\bar{\alpha}_0},\dots,\sqrt{\bar{\alpha}_T} \right \}$ in the training process then let the denoising function condition on the continuous noise level $\bar{\alpha}\sim\text{Uniform}\left (S_{t-1},S_t  \right)$. The denoising function $\boldsymbol{\epsilon}_{\theta}$ now takes three inputs with conditioned on the noisy observations $\tilde{x}$, and the loss function is formally given by:
\begin{equation}
    \mathbb{E}_{x_{0}, \bar{\alpha}, \epsilon} \left[\left\|\boldsymbol{\epsilon}-\boldsymbol{\epsilon}_{\theta}\left(\sqrt{\bar{\alpha}} x_{0}+\sqrt{1-\bar{\alpha}} \boldsymbol{\epsilon}, \tilde{x}, \bar{\alpha}\right)\right\|^{2}\right]
\end{equation}
The pseudo code for the inference phase is shown in Algorithm~\ref{alg:inference}. The inference phase in our proposed approach is straightforward to implement, which is simply applying the Eq. \eqref{eq:cond_fashion} to sample the latent variables $x_t$. To further improve the reconstruction accuracy, we take advantage of the stochasticity of the diffusion models by averaging multiple runs. This averaging process produces more precise results on the reconstructed signals (we test the 1, 3, 5, and 10-shots averages and report the results in Section~\ref{sec:results}). Recall that the DeScoD-ECG produces reconstruction of the ECG signals from a random Gaussian distribution, therefore, we could assume each reconstruction of multiple individual runs are independent to each others. Let $y$ be the clean signal and $y_m$ be the reconstruction at $m$th run, we have:
\begin{equation}
    y_{comb} = \frac{1}{M}\sum^{M}_{m=1} y_m,
\end{equation}
where $y_{comb}$ is the ensemble of multiple runs. And the output of each reconstruction can be written as the true value plus an error in the form:
\begin{equation}
     y_m = y+\varepsilon_m,
\end{equation}
and the average error made by the models acting individually (which is the 1-shot reconstruction) is therefore:
\begin{equation}
    \varepsilon_{avg}=\frac{1}{M}\sum^{M}_{m=1}(y-y_m)^2=\sum^{M}_{m=1}\frac{1}{M} \varepsilon_m^2.
\end{equation}
Similarly, the expected error from the multi-shot reconstruction is given by:
\begin{equation}
    \varepsilon_{comb}=(y-\frac{1}{M}\sum^{M}_{m=1}y_m)^2=(\sum^{M}_{m=1}\frac{1}{M} \varepsilon_m)^2.
\end{equation}
We can see obviously the multi-shots reconstruction will achieve lower error than single shot reconstruction since:
\begin{equation}
    (\sum^{M}_{m=1}\frac{1}{M} \varepsilon_m)^2 \leq \sum^{M}_{m=1}\frac{1}{M} \varepsilon_m^2.
\end{equation}

\subsection{Network Architecture} 
The architecture of the neural network used in the DeScoD-ECG is shown in Figure~\ref{fig:architecture}. The network is comprised of two stream backbones that extract features of the noisy observation $\tilde{x}$ and the latent variable $x_t$, then the Bridges to combine information of the features. The main computational module is named \emph{Half Normalized Filters} (HNF) block, which is motivated by the Multi-Kernel filter design in DeepFilter~\cite{romero2021deepfilter} and the Half-instance-norm in HINet~\cite{chen2021hinet}. The input is first fed into the multi-scale filters to extract the features under different receptive fields. Then the multi-scale features are aggregated by a convolutional layer after channel-wise concatenation. The half-instance-norm is used to normalize half of the features, which has been shown to improve the training stability and not degrade the natural statistical characters of the features~\cite{chen2021hinet}. After the half-instance-norm, the normalized and unnormalized features are fed into another convolutional layer. Then the processed features are connected with the input features with residual shortcut~\cite{he2016deep} for producing the outputs. The \emph{Bridge} block is modified based on the feature-wise linear modulation (FiLM)~\cite{dumoulin2018feature}. The bridge block directly conditions on the noise level $\sqrt{\bar{\alpha}}$ that the $\sqrt{\bar{\alpha}}$ is first encoded by sinusoidal positional embeddings~\cite{vaswani2017attention} then further encoded by a $1\times1$ convolutional layer that create the learnable scaling and shift vectors. Further, the scaling and shift vectors interact with the input features via multiplications and additions. Notably, there is no downsampling and upsampling in the whole network. Thus, the network is compatible with input signals of any length.


\begin{table}[]
\centering
\caption{ECG records used as test set. We use the exact same selections as DeepFilter~\cite{romero2021deepfilter} for fair comparison.}\label{table:list}
\begin{tabular}{|l||l|}
\hline
\textbf{Database Name}                                        & \textbf{Select ID}                          \\ \hline\hline
\multirow{2}{*}{MIT-BIH Arrhythmia Database}                  & \multirow{2}{*}{\textit{sel123, sel233}}    \\
                                                              &                                             \\ \hline
\multirow{2}{*}{MIT-BIH ST Change Database}                   & \multirow{2}{*}{\textit{sel302, sel307}}     \\
                                                              &                                             \\ \hline
\multirow{2}{*}{MIT-BIH Supraventricular Arrhythmia Database} & \multirow{2}{*}{\textit{sel820, sel853}}     \\
                                                              &                                             \\ \hline
\multirow{2}{*}{MIT-BIH Normal Sinus Rhythm Database}         & \multirow{2}{*}{\textit{sel16420, sel16795}} \\
                                                              &                                             \\ \hline
\multirow{2}{*}{European ST-T Database}                       & \multirow{2}{*}{\textit{sel0106, sel0121}}   \\
                                                              &                                             \\ \hline
\multirow{2}{*}{Sudden death patients from BIH}               & \multirow{2}{*}{\textit{sel32, sel49}}       \\
                                                              &                                             \\ \hline
\multirow{2}{*}{MIT-BIH LongTerm ECG Database}                & \multirow{2}{*}{\textit{sel14046, sel15815}} \\
                                                              &                                             \\ \hline
\end{tabular}
\end{table}

\begin{figure*}
    \centering
    \subfigure[Time domain of a segment from \textit{sel123}]{
        \includegraphics[width=.45\textwidth]{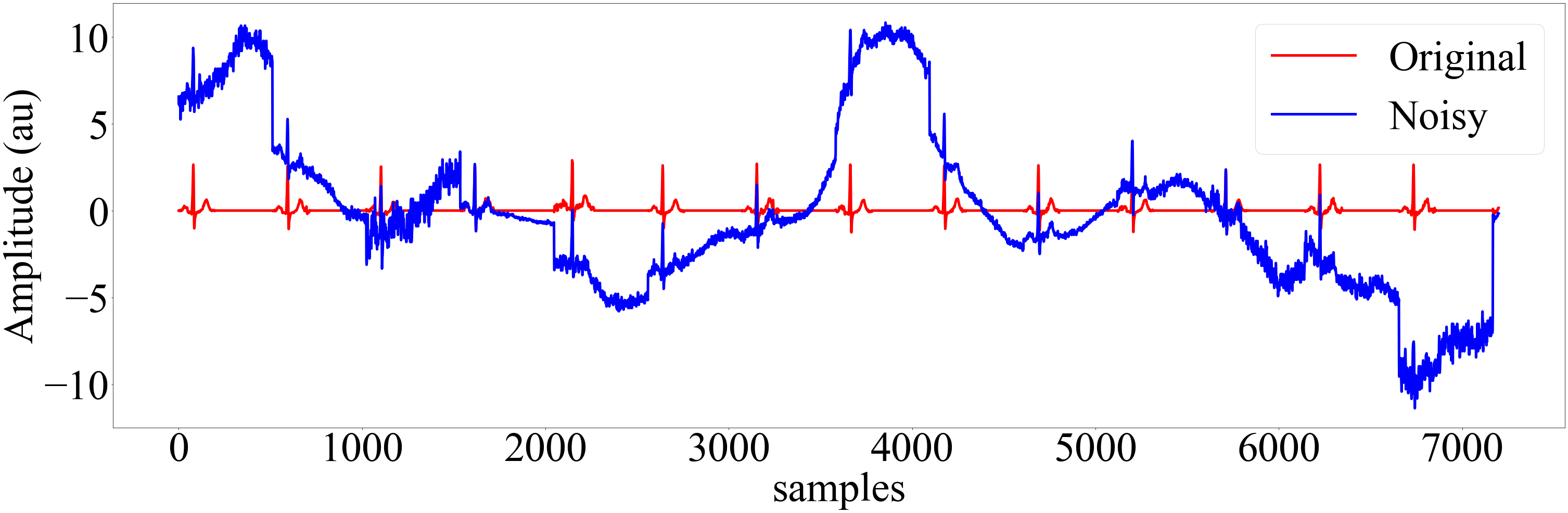}
    }
    \subfigure[Power spectral density of the segment from \textit{sel123}]{
        \includegraphics[width=.45\textwidth]{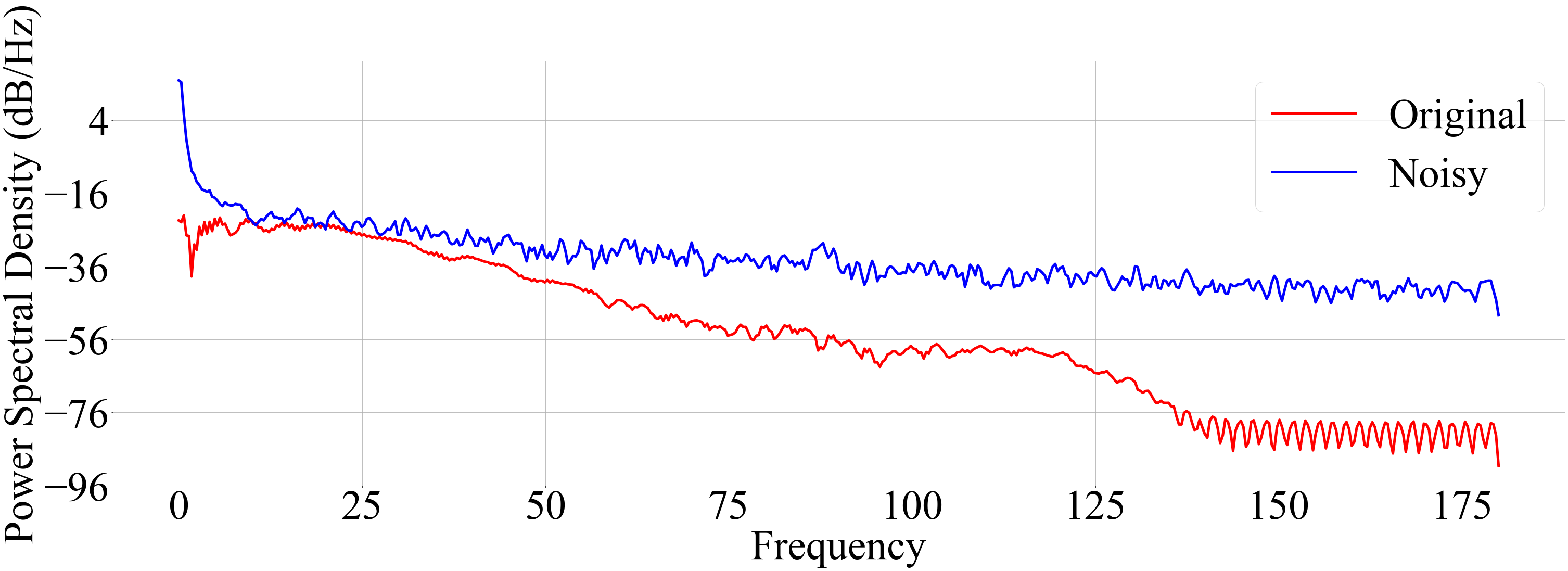}
    }
    \subfigure[Time domain of a segment from \textit{sel302}]{
        \includegraphics[width=.45\textwidth]{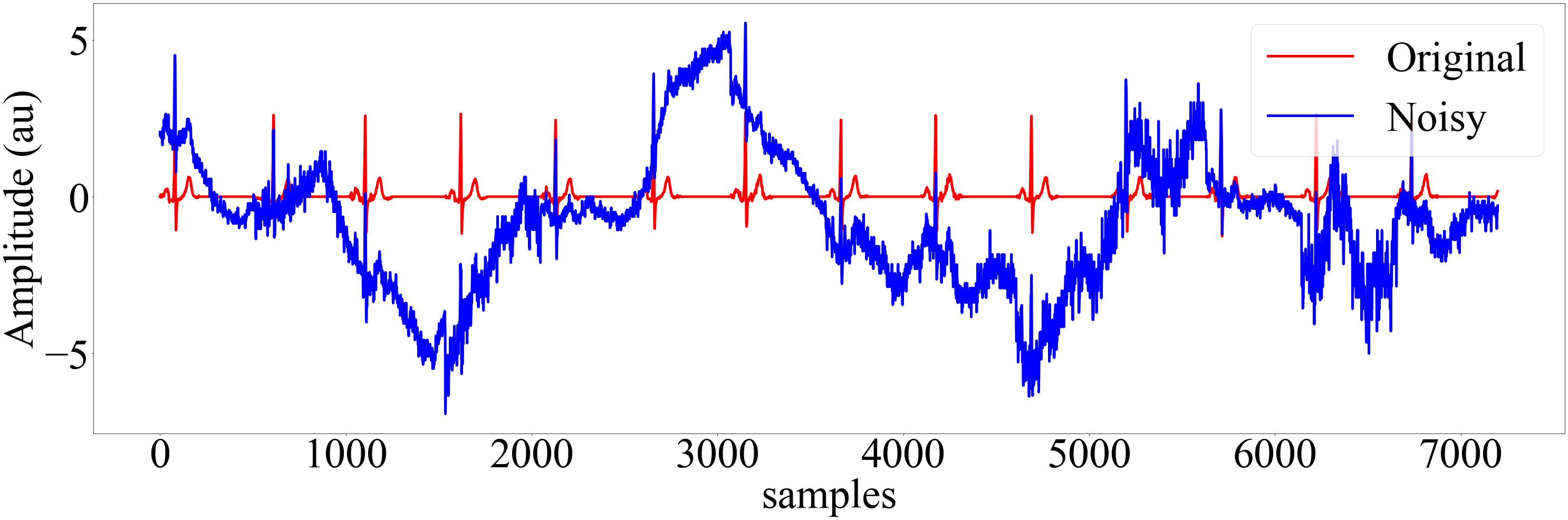}
    }
    \subfigure[Power spectral density of the segment from \textit{sel302}]{
        \includegraphics[width=.45\textwidth]{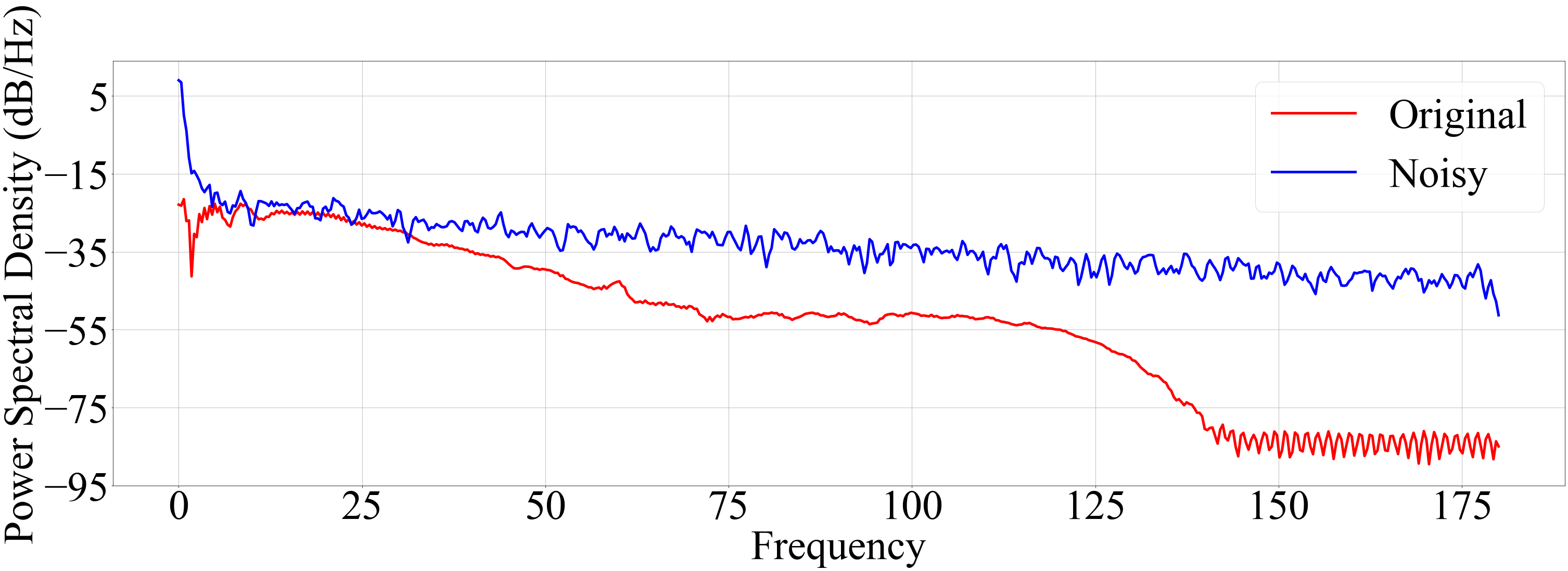}
    }
    \caption{Visualization of the clean (red lines) and noisy (blue lines) ECG records in the time domain and the corresponding power spectral density plot.}
    \label{fig:psd}
\end{figure*}

\begin{table*}
\centering
\caption{Evaluation metrics results for different methods for random noise amplitude between 0.2 to 2. Comparison results are obtained from~\cite{romero2021deepfilter}.}
\begin{tabular}{ccccc}
\hline\hline
Method/Model   & SSD (au)                   & MAD (au)                   & PRD (\%)                     & Cosine Sim                \\ \hline
FIR filter     & 47.464 $\pm$ 88.496        & 0.676 $\pm$ 0.572          & 66.158 $\pm$ 22.104          & 0.692 $\pm$ 0.216         \\\hline
IIR filter     & 35.645 $\pm$ 70.189        & 0.594 $\pm$ 0.532          & 61.125 $\pm$ 23.220          & 0.736 $\pm$ 0.208         \\\hline
FCN-DAE        & 10.633 $\pm$ 14.417        & 0.598 $\pm$ 0.381          & 94.504 $\pm$ 70.717          & 0.770 $\pm$ 0.171         \\\hline
CGAN           & 7.834 $\pm$ 13.224         & 0.442 $\pm$ 0.323          & 57.133 $\pm$ 41.816          & 0.852 $\pm$ 0.141         \\\hline
DRNN           & 6.702 $\pm$ 11.316         & 0.480 $\pm$ 0.340          & 50.035 $\pm$ 24.802          & 0.878 $\pm$ 0.126         \\\hline
DeepFilter     & 5.203 $\pm$ 7.915          & 0.386 $\pm$ 0.283          & 50.445 $\pm$ 29.598          & 0.897 $\pm$ 0.097         \\\hline
Ours (1-shot)  & 4.912 $\pm$ 7.822          & 0.369 $\pm$ 0.289          & 44.218 $\pm$ 27.980          & 0.904$\pm$ 0.113          \\\hline
Ours (3-shot)  & 4.057$\pm$ 6.202           & 0.340 $\pm$ 0.266          & 41.523 $\pm$ 26.169          & 0.920 $\pm$ 0.095         \\\hline
Ours (5-shot)  & 3.887 $\pm$ 5.911          & 0.334 $\pm$ 0.261          & 40.987$\pm$ 26.562           & 0.923 $\pm$ 0.091         \\ \hline
Ours (10-shot) & \textbf{3.771 $\pm$ 5.713} & \textbf{0.329 $\pm$ 0.258} & \textbf{40.527 $\pm$ 26.258} & \textbf{0.926$\pm$ 0.087} \\\hline
\end{tabular}
\label{tabel:bw}
\end{table*}

\begin{table*}[]
\centering
\caption{SSD in different noise segments.}
\begin{tabular}{ccccc}
\hline\hline
Method/Model    & $0.2$ to $0.6$     & $0.6$ to $1.0$      & $1.0$ to $1.5$      & $1.5$ to $2.0$   \\ \hline
FIR filter     & 7.162 $\pm$ 9.485          & 22.542 $\pm$ 30.795        & 51.381 $\pm$ 72.620        & 95.983 $\pm$ 134.621       \\
IIR filter     & 5.615 $\pm$ 7.508          & 17.262 $\pm$ 25.049        & 38.334 $\pm$ 58.050        & 71.980 $\pm$ 107.880       \\
FCN-DAE        & 7.396 $\pm$ 8.032          & 8.638 $\pm$ 8.247          & 10.929 $\pm$ 12.883        & 14.586 $\pm$ 21.429        \\
DRNN           & 4.000 $\pm$ 7.437          & 4.932 $\pm$ 6.652          & 7.016 $\pm$ 10.971         & 10.031 $\pm$ 15.665        \\
CGAN           & 3.461 $\pm$ 6.054          & 5.518 $\pm$ 7.994          & 8.579 $\pm$ 12.565         & 12.382 $\pm$ 18.598       \\
DeepFilter     & 2.401 $\pm$ 3.419          & 3.816 $\pm$ 4.771          & 5.717 $\pm$ 7.781          & 8.078 $\pm$ 11.104         \\\hline
Ours (1-shot)  & 2.324 $\pm$ 3.798          & 3.762 $\pm$ 5.869          & 5.328 $\pm$ 7.195          & 7.466 $\pm$ 10.724         \\
Ours (3-shot)  & 2.036 $\pm$ 3.225          & 3.252 $\pm$ 5.261          & 4.421 $\pm$ 5.739          & 5.935 $\pm$ 8.164          \\
Ours (5-shot)  & 1.977 $\pm$ 3.201          & 3.147 $\pm$ 4.963          & 4.193 $\pm$ 5.474          & 5.686 $\pm$ 7.772          \\
Ours (10-shot) & \textbf{1.953 $\pm$ 3.138} & \textbf{3.046 $\pm$ 4.902} & \textbf{4.081 $\pm$ 5.315} & \textbf{5.483 $\pm$ 7.433}\\ \hline
\end{tabular}
\label{tabel:bw_ssd}
\end{table*}

\begin{table*}[]
\centering
\caption{MAD in different noise segments.}
\begin{tabular}{ccccc}
\hline\hline
Method/Model   & $0.2$ to $0.6$     & $0.6$ to $1.0$     & $1.0$ to $1.5$     & $1.5$ to $2.0$       \\ \hline
FIR filter     & 0.280 $\pm$ 0.183          & 0.510 $\pm$ 0.337          & 0.759 $\pm$ 0.513          & 1.045 $\pm$ 0.713          \\
IIR filter     & 0.248 $\pm$ 0.171          & 0.449 $\pm$ 0.322          & 0.666 $\pm$ 0.487          & 0.918 $\pm$ 0.676          \\
FCN-DAE        & 0.480 $\pm$ 0.278          & 0.532 $\pm$ 0.301          & 0.617 $\pm$ 0.369          & 0.727 $\pm$ 0.474          \\
DRNN           & 0.390 $\pm$ 0.286          & 0.432 $\pm$ 0.273          & 0.496 $\pm$ 0.330          & 0.576 $\pm$ 0.409          \\
CGAN           & 0.297 $\pm$ 0.175          & 0.370 $\pm$ 0.215          & 0.468 $\pm$ 0.306          & 0.587 $\pm$ 0.419          \\
DeepFilter     & 0.293 $\pm$ 0.214          & 0.341 $\pm$ 0.222          & 0.401 $\pm$ 0.270          & 0.483 $\pm$ 0.351          \\\hline
Ours (1-shot)  & 0.230 $\pm$ 0.188          & 0.321 $\pm$ 0.225          & 0.403 $\pm$ 0.275          & 0.484 $\pm$ 0.351          \\
Ours (3-shot)  & 0.216 $\pm$ 0.180          & 0.299 $\pm$ 0.213          & 0.369 $\pm$ 0.251          & 0.442 $\pm$ 0.323          \\
Ours (5-shot)  & 0.213 $\pm$ 0.178          & 0.295 $\pm$ 0.211          & 0.362 $\pm$ 0.247          & 0.432 $\pm$ 0.314          \\
Ours (10-shot) & \textbf{0.211 $\pm$ 0.178} & \textbf{0.291 $\pm$ 0.210} & \textbf{0.356 $\pm$ 0.243} & \textbf{0.426 $\pm$ 0.313} \\ \hline
\end{tabular}
\label{tabel:bw_mad}
\end{table*}

\begin{table*}[]
\centering
\caption{PRD in different noise segments.}
\begin{tabular}{ccccc}
\hline\hline
Method/Model    & $0.2$ to $0.6$     & $0.6$ to $1.0$      & $1.0$ to $1.5$       & $1.5$ to $2.0$         \\ \hline
FIR filter     & 42.673 $\pm$ 15.504          & 61.370 $\pm$ 18.213          & 73.491 $\pm$ 17.577          & 81.627 $\pm$ 15.250          \\
IIR filter     & 38.653 $\pm$ 15.668          & 55.867 $\pm$ 19.482          & 67.972 $\pm$ 19.714          & 76.644 $\pm$ 18.290          \\
FCN-DAE        & 76.445 $\pm$ 41.696          & 86.350 $\pm$ 56.163          & 98.331 $\pm$ 68.010          & 112.012 $\pm$ 94.656         \\
DRNN           & 40.298 $\pm$ 19.983          & 45.582 $\pm$ 21.936          & 52.317 $\pm$ 24.336          & 59.312 $\pm$ 27.389          \\
CGAN           & 39.090 $\pm$ 19.669          & 48.972 $\pm$ 24.694          & 60.514 $\pm$ 36.254          & 74.830 $\pm$ 59.722          \\
DeepFilter     & 34.026 $\pm$ 14.234          & 43.346 $\pm$ 19.925          & 53.777 $\pm$ 27.189          & 66.167 $\pm$ 38.040          \\\hline
Ours (1-shot)  & 29.762 $\pm$ 15.583          & 38.945 $\pm$ 20.183          & 47.678 $\pm$ 25.366          & 56.559 $\pm$ 35.969          \\
Ours (3-shot)  & 28.066 $\pm$ 14.923          & 36.681 $\pm$ 19.165          & 44.506 $\pm$ 23.090          & 53.225 $\pm$ 33.843          \\
Ours (5-shot)  & 27.697 $\pm$ 14.706          & 36.187 $\pm$ 18.974          & 43.774 $\pm$ 22.834          & 52.715 $\pm$ 35.341          \\
Ours (10-shot) & \textbf{27.475 $\pm$ 14.631} & \textbf{35.710 $\pm$ 18.718} & \textbf{43.321 $\pm$ 22.959} & \textbf{52.081 $\pm$ 34.681} \\ \hline
\end{tabular}
\label{tabel:bw_prd}
\end{table*}

\begin{table*}[]
\centering
\caption{Cosine similarities in different noise segments.}
\begin{tabular}{ccccc}
\hline\hline
Method/Model   & $0.2$ to $0.6$    & $0.6$ to $1.0$     & $1.0$ to $1.5$      & $1.5$ to $2.0$   \\\hline
FIR filter     & 0.891 $\pm$ 0.082          & 0.760 $\pm$ 0.149          & 0.637 $\pm$ 0.192          & 0.532 $\pm$ 0.209          \\
IIR filter     & 0.910 $\pm$ 0.074          & 0.800 $\pm$ 0.143          & 0.690 $\pm$ 0.191          & 0.589 $\pm$ 0.217          \\
FCN-DAE        & 0.844 $\pm$ 0.110          & 0.807 $\pm$ 0.136          & 0.757 $\pm$ 0.167          & 0.692 $\pm$ 0.205          \\
DRNN           & 0.926 $\pm$ 0.070          & 0.906 $\pm$ 0.084          & 0.871 $\pm$ 0.122          & 0.824 $\pm$ 0.167          \\
CGAN           & 0.928 $\pm$ 0.060          & 0.891 $\pm$ 0.085          & 0.838 $\pm$ 0.131          & 0.773 $\pm$ 0.184          \\
DeepFilter     & 0.948 $\pm$ 0.042          & 0.921 $\pm$ 0.064          & 0.888 $\pm$ 0.093          & 0.844 $\pm$ 0.124          \\\hline
Ours (1-shot)  & 0.955 $\pm$ 0.049          & 0.929 $\pm$ 0.068          & 0.895 $\pm$ 0.104          & 0.853 $\pm$ 0.156          \\
Ours (3-shot)  & 0.961 $\pm$ 0.042          & 0.939 $\pm$ 0.058          & 0.913 $\pm$ 0.083          & 0.879 $\pm$ 0.132          \\
Ours (5-shot)  & 0.961 $\pm$ 0.041          & 0.941 $\pm$ 0.056          & 0.918 $\pm$ 0.078          & 0.883 $\pm$ 0.129          \\
Ours (10-shot) & \textbf{0.962 $\pm$ 0.040} & \textbf{0.943 $\pm$ 0.055} & \textbf{0.920 $\pm$ 0.077} & \textbf{0.889 $\pm$ 0.124} \\\hline
\end{tabular}
\label{tabel:bw_cos}
\end{table*}

\begin{figure*}
    \centering
    \subfigure[0.2 $<$ noise $<$ 0.6]{
        \includegraphics[width=.85\textwidth]{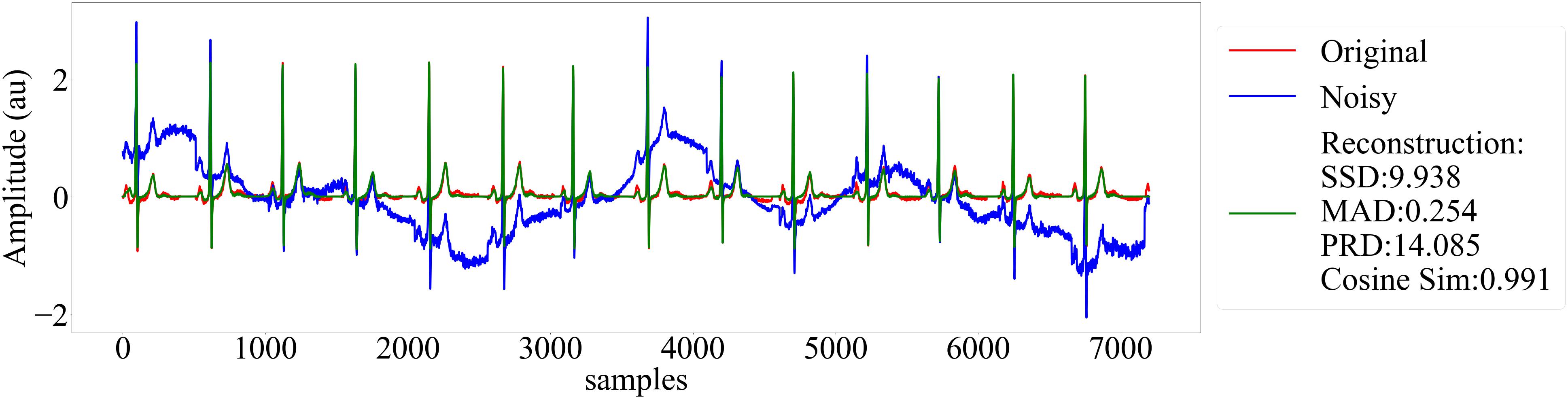}
    }
    \subfigure[0.6 $<$ noise $<$ 1.0]{
        \includegraphics[width=.85\textwidth]{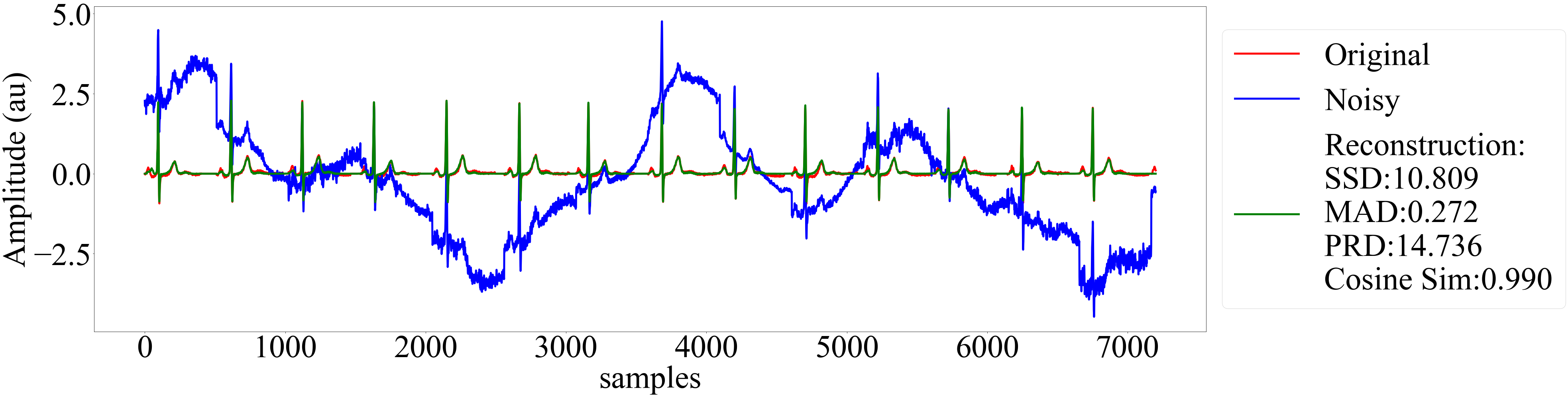}
    }
    \subfigure[1.0 $<$ noise $<$ 1.5]{
        \includegraphics[width=.85\textwidth]{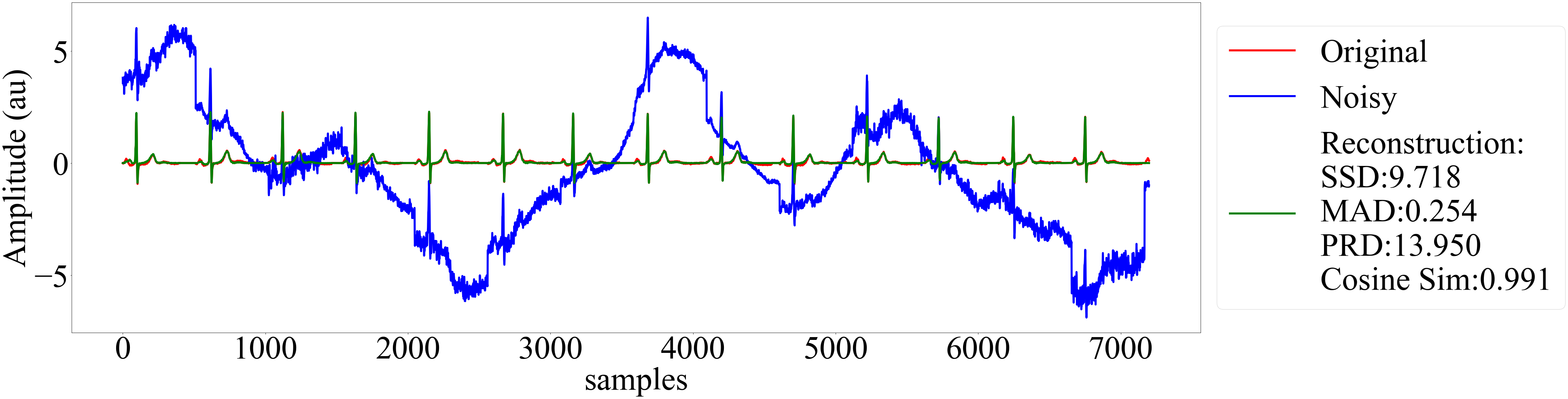}
    }
    \subfigure[1.5 $<$ noise $<$ 2.0]{
        \includegraphics[width=.85\textwidth]{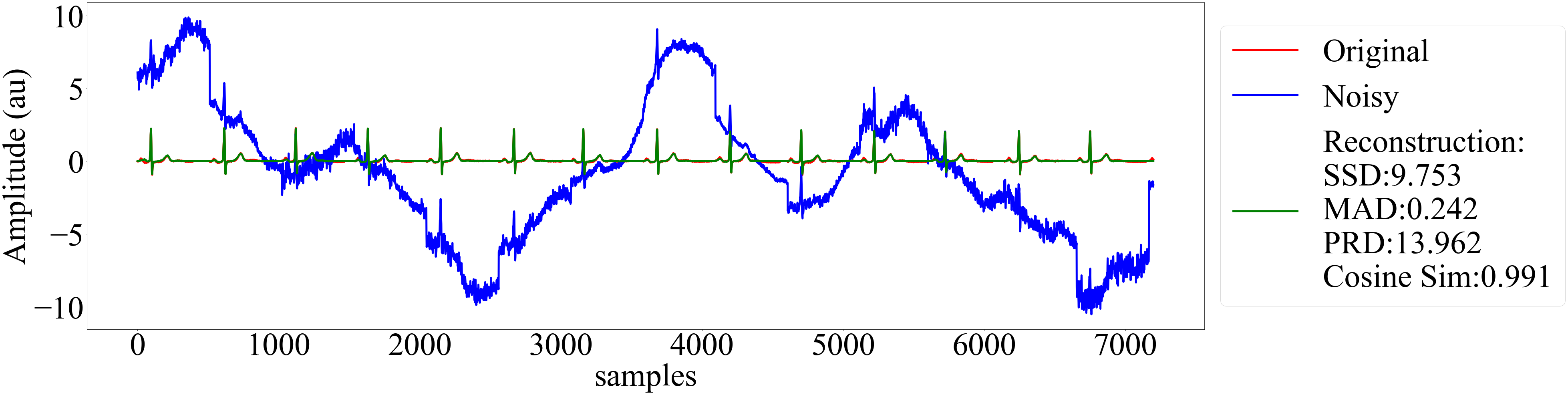}
    }
    \caption{The reconstructions of a segment intercepted from the \textit{sele123} records under different noise amplitudes.}
    \label{fig:examples}
\end{figure*}

\section{Experiments}
In this section, we give the details of the experiments. We first introduce the definitions and formulations of the four similarity metrics that quantify the quality of the reconstructed ECG signals. Then we summarize the datasets and introduce our experimental procedure. Finally, we introduce the baseline comparison methods and discuss the quantitative evaluations of the DeScoD-ECGs and the baselines. We also visualize reconstruction results and how the signals change during the reverse process.

\subsection{Evaluation Metrics}


This work evaluates the distortion of the ECG signals in a quantitative way with four distance-based metrics as suggested by multiple previous studies~\cite{nygaard2001rate, manikandan2008ecg,romero2018baseline}. According to those studies, distance-based metrics are considered more suitable than statistics-based metrics (such as Root Mean square error (RMSE), Cross-correlation (CC), and Signal noise
ratio (SNR)) for calculating the dissimilarity/similarity between the clean and restored signals. For convenience, we use the notations $x$, $\tilde{x}$, and $\hat{x}$ to refer to the clean, the noisy, and denoised ECGs, respectively. The definitions and formulations of the four metrics are as follows:

\begin{itemize}
\item \textbf{Sum of the square of the distances (SSD)~\cite{nygaard2001rate}}: SSD is a variant of RMSE which is defined as:
\begin{equation}
    \text{SSD} = \sum^{N-1}_{n=0}\left [x(n)-\hat{x}(n)\right ]^2.
\end{equation}
Unlike RMSE, the SSD of a reconstructed signal provides an idea of how similar the signals are along their entire duration instead of average difference.

\item \textbf{Absolute maximum distance (MAD)~\cite{nygaard2001rate}}:  MAD is a well-known similarity metric to quantify the ECG quality. Specifically, MAD measures the maximum absolute distance between the original and filtered signals. MAD is given by:
\begin{equation}
    \text{MAD} = \max\left | x(n)-\hat{x}(n) \right |, \quad \text{for} \quad 0\leq n\leq N .
\end{equation}

\item \textbf{Percentage root-mean-square difference (PRD)~\cite{nygaard2001rate}}:  PRD computes the total distortion present in the denoised signal. The calculation of PRD is given by:
\begin{equation}
    \text{PRD} = \sqrt{\frac{\sum^{N-1}_{n=0}\left [x(n)-\hat{x}(n)\right ]^2}{\sum^{N-1}_{n=0}\left [x(n)-\frac{1}{N}\sum^{N-1}_{n=0}x(n)\right ]^2}}\times 100\%.
\end{equation}

\item \textbf{Cosine similarity (Cosine Sim)~\cite{nygaard2001rate}}: The cosine similarity is a measure of similarity between two vectors through the normalized dot product by the Euclidean L2 normalization. The cosine similarity is given by:
\begin{equation}
    \text{Cosine Sim} = \frac{\left \langle x, \hat{x} \right \rangle}{\left \| x \right \|\left \| \hat{x} \right \|}.
\end{equation}

\end{itemize}

\subsection{Dataset}

We benchmark the DeScoD-ECG on the QT Database~\cite{laguna1997database} and the Massachusetts Institute of Technology -- Beth Israel Hospital Noise Stress Test Database (MIT-BIH NST)~\cite{Moody84}. The QT Database and MIT-BIH Noise Stress Test Database are archived and available at \url{www.physionet.org}~\cite{physiobank2000physionet}. We use the ECG records in the QT database, and corrupt the ECG records with the noise profiles taken from the MIT-BIH NST. QT Database is formed of 105 15-min two-lead ECG recordings containing a wide range of QRS and ST-T morphologies in all the possible variability. The MIT-BIH NST contains three noise recordings of typical noise that appears in stress tests which are caused by motion-related interference. The electrodes are placed on the patients' limbs to collect the noises without the presence of ECG signals. 

When adding the noise to the ECG records, we first normalized the noise records to the range of min and max value of the corresponding ECG signals and rescaled the noise amplitudes by multiplying a random factor between 0.2 to 2.0. By doing that, we have a better chance to observe the performance of the denoising methods under different noisy conditions. Figure. \ref{fig:psd} visualized some examples of the ECG signals with noise in both the time and frequency domains. We can observe that the baseline wander noises used in this work are more than just low-frequency noise that warps the signals. The noises contain both low and high-frequency components that can be used to prove the generalization capability of the denoising algorithms.

\subsection{Training and implementation details}
We use the 14 signals listed in Table~\ref{table:list} as a test set, which are the same as the experiments in DeepFilter~\cite{romero2021deepfilter} for a fair comparison. Also, selecting test signals from different sources than the training set can ensure the generalization of the algorithms. The remaining ECG records are used for training and validation. We randomly shuffle the remaining records and split them by 70\% for training and 30\% for validation. For the convenience of training and testing, we split the signals into small patches with 512 points. Therefore, we obtained 50723 samples as training data, 21738 samples as the validation set, and 13337 as the test set.
The random seed was set to the one used in the DeepFilter experiments to add noise to the original ECG signal. We ran an experiment to verify this was the case by re-running the FIR and IIR experiments and obtaining the same results. 

We implemented the proposed using the PyTorch framework~\cite{paszke2019pytorch}, and the source code will be freely available if this work is accepted\footnote{\url{https://github.com/HuayuLiArizona/Score-based-ECG-Denoising.git}}. 
During training, a batch size of 96 is used. We trained the model for 400 epochs by Adam optimizer~\cite{kingma2014adam} with an initial learning rate of 0.001, and the learning rate decays by multiplying 0.1 every 150 epochs. We save the model weights with the lowest loss on the validation set during training and evaluate the performance only on the validation set for model selection. That means, before we determine the final model for testing, the test set is invisible to the model. We do that because we do not want to adjust the network architecture and the hyper-parameters to overfit the test set.

\subsection{Results}
\label{sec:results}

\begin{figure*}
    \centering
    \subfigure[Iteration 0.]{
        \includegraphics[width=.31\textwidth]{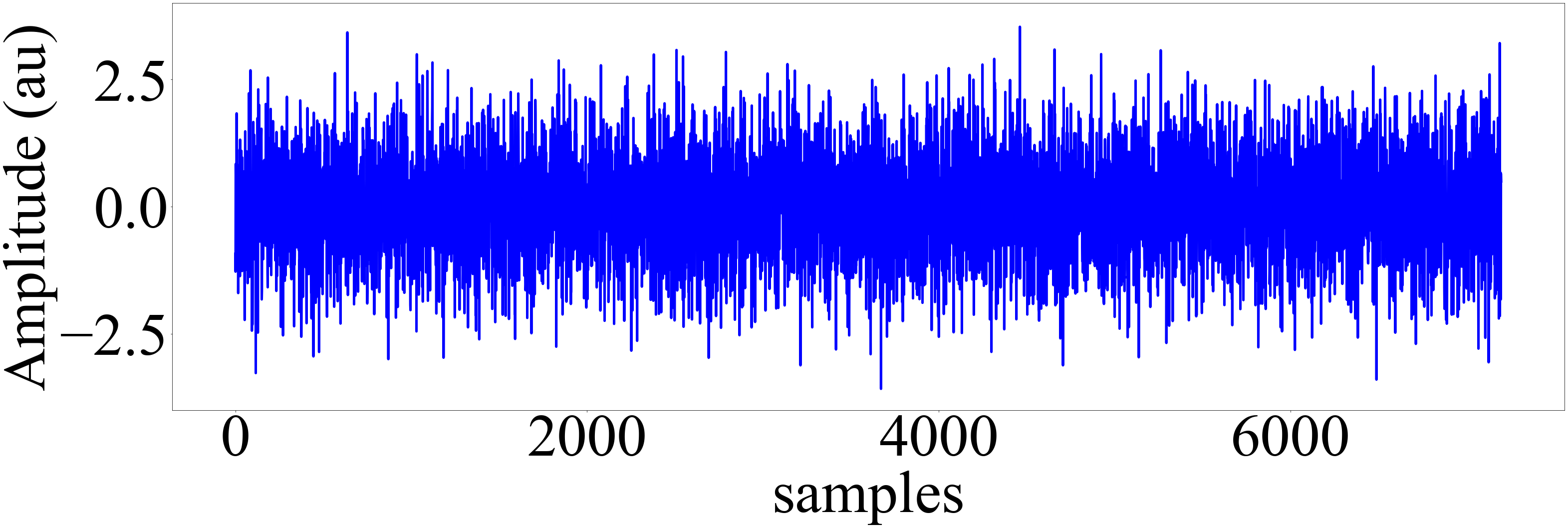}
    }
    \subfigure[Iteration 10.]{
        \includegraphics[width=.31\textwidth]{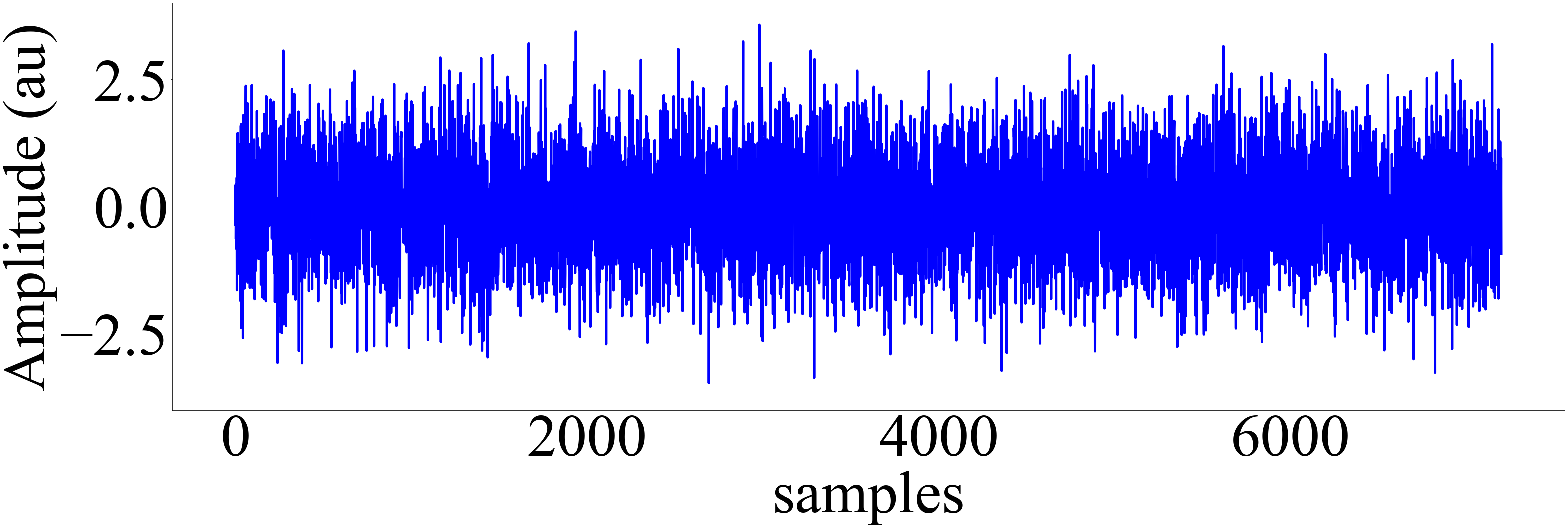}
    }
    \subfigure[Iteration 20.]{
        \includegraphics[width=.31\textwidth]{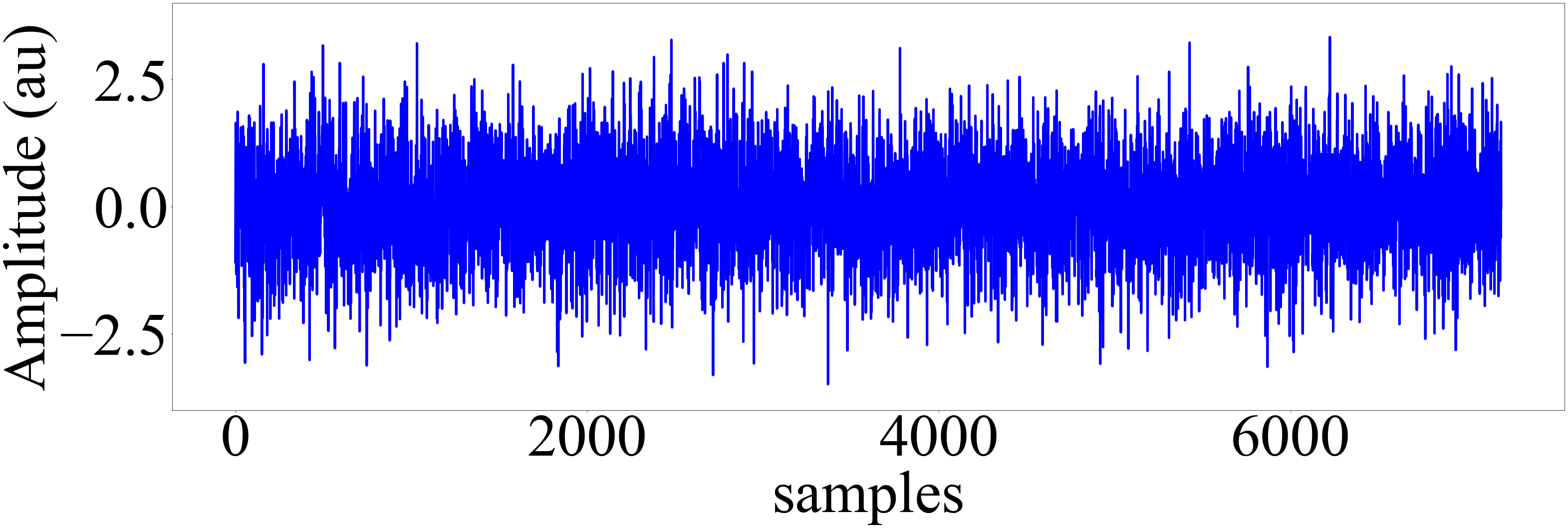}
    }
    \subfigure[Iteration 30.]{
        \includegraphics[width=.31\textwidth]{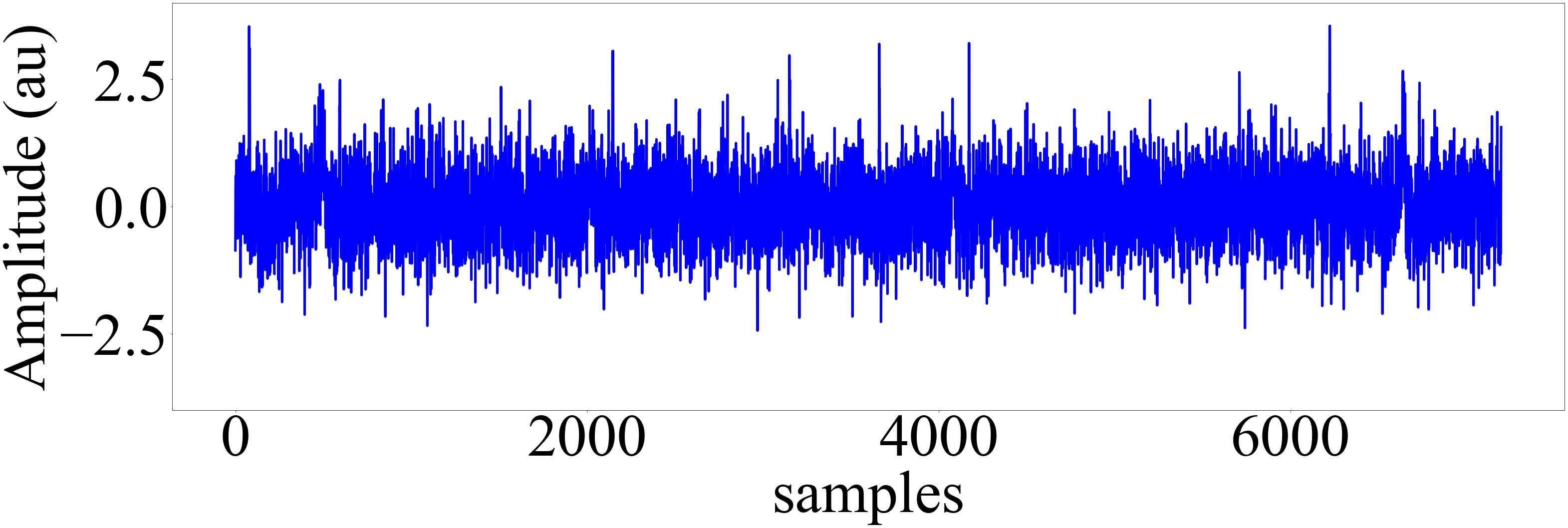}
    }
    \subfigure[Iteration 40.]{
        \includegraphics[width=.31\textwidth]{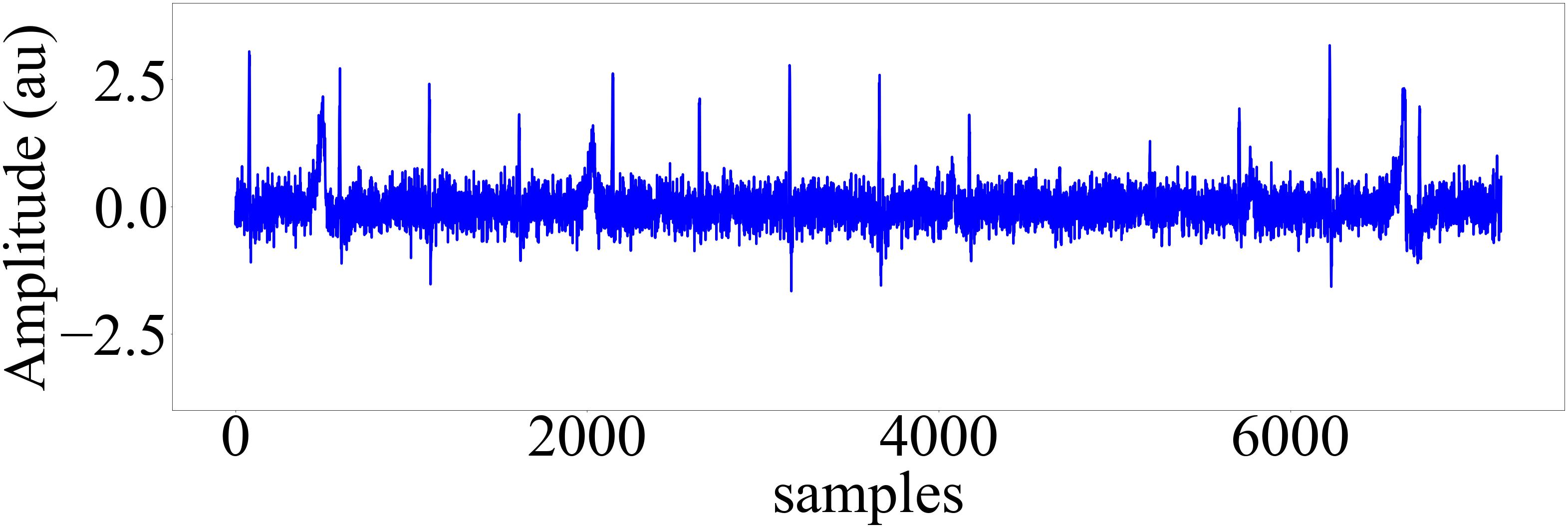}
    }
    \subfigure[Iteration 50.]{
        \includegraphics[width=.31\textwidth]{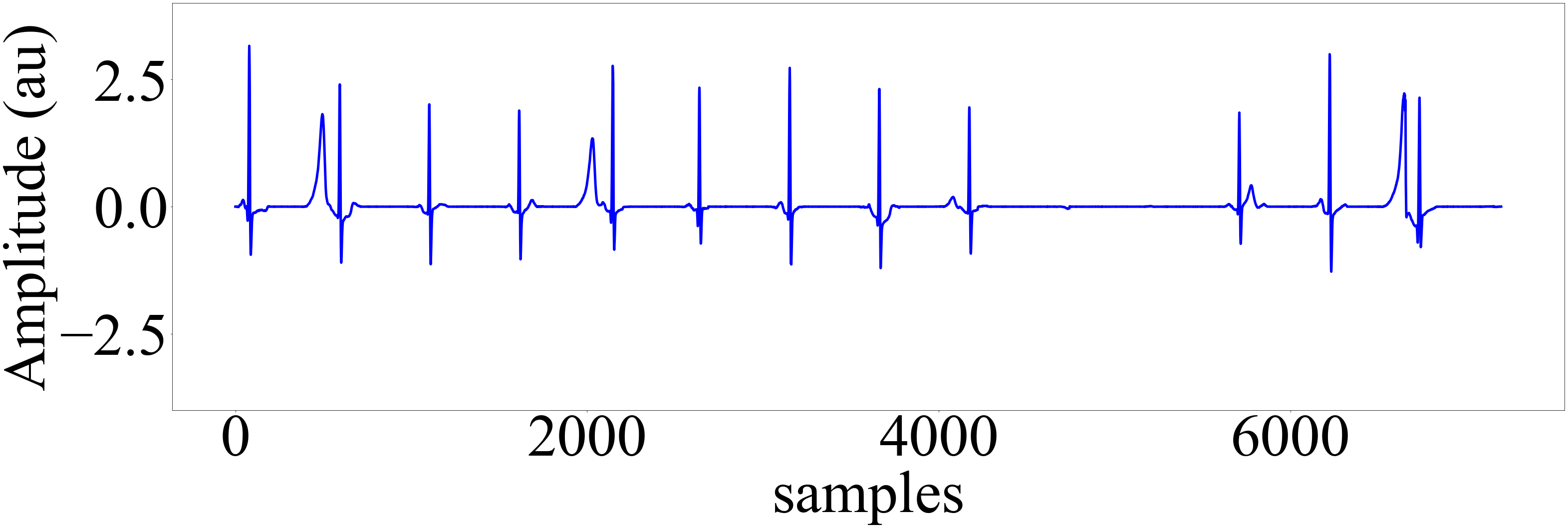}
    }
    \caption{Visualization of the inference process which the diffusion model denoise the Gaussian noise iteratively to the approximation of true data.}
    \label{fig:iters}
\end{figure*}

We compare the performance of the DeScoD-ECG with several representative baseline methods, namely: FIR and IIR filters, DRNN~\cite{antczak2018deep}, FCN-DAE~\cite{chiang2019noise}, CGAN~\cite{wang2022ecg}, and DeepFilter~\cite{romero2021deepfilter}. We present the comparisons in Table~\ref{tabel:bw} which are shown as the mean values and standard deviations of the metrics. We can observe that the DeScoD-ECG achieves better performance than the baselines even with 1-shot reconstruction. With 1-shot reconstruction, the DeScoD-ECG outperforms the comparative methods in all the metrics. That indicates the score-based diffusion model achieved better results than na\"ively end-to-end training neural networks with better distribution approximation capability. When we use multi-shot reconstruction, the performance is further improved. For example, the 10-shot reconstruction obtained the best results on all four similarity metrics with 3.771 $\pm$ 5.713 au, 0.329 $\pm$ 0.258 au, 40.527 $\pm$ 26.258 \%, and 0.926 $\pm$ 0.087 compared to the existing state-of-the-art method, DeepFilter, which achieved 5.203 $\pm$ 7.915 au, 0.386 $\pm$ 0.283 au, 50.455 $\pm$ 29.598 \%, and, 0.897 $\pm$ 0.097. The quantitative evaluations on multi-shot reconstruction support our hypothesis that the DeScoD-ECG can produce more accurate results with a self-ensemble strategy, which averages the multiple predictions.

Further, we evaluate the model performance on the test set under different noise amplitudes. Table \ref{tabel:bw_ssd} to \ref{tabel:bw_cos} present the four metrics of different noise segments. The results on each noise segment and different shots are consistent with the overall performance in Table \ref{tabel:bw}. We notice several observations that could further prove the efficiency of the DeScoD-ECGs. First, the DeScoD-ECG with 1-shot reconstruction still outperforms the baselines and achieves better results than the existing methods in 16 out of 16 evaluations under different noise segments. Second, the gain brought by multi-shots reconstruction still holds under different noise strengths. Third, the DeScoD-ECG boosted by the multi-shots reconstruction is more stable under extreme noisy scenarios than the baselines. For example, the results of 10-shot reconstruction under noise amplitude between 1.5 and 2.0 outperforms the results of DeepFilter under noise amplitude between 1.0 and 1.5. Figure \ref{fig:examples} shows the reconstruction results of an ECG beat extracted from the signal \textit{sel123} under different noise amplitude. We observed that the DeScoD-ECG could produce precise and consistent reconstruction under different noise levels and keep the morphology of the signal with a significant baseline wander. 

Another interesting observation is how the DeScoD-ECG reconstructs the signal starting with random Gaussian noise. A sequence of results sampled from the reverse process is visualized in Figure \ref{fig:iters}. We can see that the diffusion model denoises the Gaussian noise iteratively by a small step. After 50 iterations, the Gaussian noise is denoised and transformed into a clean ECG signal.

\begin{table}[]
\centering
\caption{Numbers of parameters of different models.}
\begin{tabular}{|c|c|c|c|c|c|}
\hline
Method     & FCN-DAE & DRNN   & CGAN   & DeepFilter & Ours \\ \hline
Parameters & 0.08 M  & 0.03 M & 4.89 M & 0.17 M     & 1.22 M     \\ \hline
\end{tabular}
\label{tabel:para}
\end{table}

\section{Discussion and Future works}
This work proposed DeScoD-ECG, a deep score-based diffusion model for ECG baseline wander and noise removal. Experiments conducted on the QT database and MIT-BIH NST demonstrate the facts in two aspects. First, we show that the DeScoD-ECG achieved the best performance compared to the state-of-the-art denoising solutions. Taking advantage of the excellence of score-based diffusion models in approximating the true data distribution, DeScoD-ECG is proven able to beat the baseline methods with high accuracy and the preservation of fine details. The results reported in Table \ref{tabel:bw} show that DeScoD-ECG has the best overall denoising performance in four distance-based metrics. The results reported in Table \ref{tabel:bw_ssd} to \ref{tabel:bw_cos} show the denoising performance under different noise levels is consistent with overall performance and also surpasses the baseline methods. 
Second, we demonstrate that the self-ensembles strategies can further boost the precision of the reconstructions. The DeScoD-ECG also shows upward trends in the general performance and the performance by noise level segments for the four metrics when increasing ensemble size. 
In summary, the first fact proves the outstanding performance of the score-based diffusion model, and the second shows the existence of strategies for further improving the score-based diffusion model. 

The DeScoD-ECG is of great significance for healthcare research. The motion-related noises, including electrode motion artifacts and baseline wander, exist widely in wearable ECG monitoring devices~\cite{webster2009medical, an2020comparison}. Therefore, the noises used in this work are rational to prove the feasibility of applying the DeScoD-ECG in filtering the ECG collected in exercise stress tests. The excellent performance of the DeScoD-ECG under extreme noisy environments enables the exercise stress tests to take place in more flexible scenarios. This work can be regarded as a preliminary study and a critical data processing tool for our future analysis of exercise stress tests beyond the traditional test items, such as walking on a treadmill or riding a stationary bike. The routine examination items limit the patients' body movement, activity space, and exercise intensity to prevent the influence of the noises. However, the applications of stable and accurate reconstruction of the noisy ECG allow us conducted stress tests in more flexible ways. For instance, the impacts of COVID-19 on elite athletes are studied in \cite{yeo2020sport}. Athletes risk persistent cardiac or incapacitating symptoms that stop them from resuming competitive sports. ECG monitoring in the exercise stress test is critical for the cardiovascular evaluation of athletes. In particular, these stress tests are vital to evaluate whether an athlete's health condition allows them to return to competition. 
However, their hidden cardiovascular dangers may only be exposed when they go through their own routine training~\cite{sridi2022resuming}. Therefore, we need ECG denoising techniques that can perform well in extreme noise conditions to fulfill the flexible exercise stress test requirements. 

In addition to the performance achieved by DeScoD-ECG, we find some interesting future work directions. One direction is to improve the computation efficiency. We evaluated the running speed of DeScoD-ECG on a regular configuration laptop with i7-10750H 2.60GHz CPU and GTX 1660 Ti GPU. We chose this configuration due to the consideration of verifying DeScoD-ECG could be run by affordable personal computers for researchers who want to use DeScoD-ECG for ECG denoising. Also, we provided the numbers of parameter of DeScoD-ECG and the baseline methods in Tabel \ref{tabel:para}. We test the running time of the DeScoD-ECG on a 20-second ECG record (7200 points with a sample rate of 360 Hz) that the 1, 3, 5, and 10-shot reconstructions take 0.682, 1.970, 3.284, and 7.136 second. Although the running time is shorter than the record duration, and the running speed can be further improved by batch computation, we should think about a fast version of DeScoD-ECG in our future works. The major computational overhead of diffusion models is brought by the sampling/inference process. Thus, speeding up the sampling efficiency of the diffusion model is of great significance in improving DeScoD-ECG. Several recent studies~\cite{song2020score,kong2020diffwave} were conducted to accelerate the reversing process of diffusion models by leveraging ODE solvers. Thus, our next step will focus on combining the DeScoD-ECG with ODE solvers to mitigate the computational overhead. Another direction is to enhance the self-ensemble strategy. The DeScoD-ECG used a simple averaging over the multi-shots reconstruction, while we believe that developing a weighted ensemble strategy will further improve the quality of the restored signals. Therefore, in our future works, we will explore the possibility of using quality assignment techniques~\cite{satija2018review} to give each reconstruction a weight factor to achieve the weighted self-ensemble strategy. Also, we will evaluate the DeScoD-ECG on more types of noise and deploy the denoised results to more downstream tasks.

\section{Conclusion}
In this paper, we presented DeScoD-ECG, a novel ECG baseline wander and noise removal approach based on a conditional score-based diffusion model, which has achieved state-of-the-art performance compared with existing baseline methods. Unlike the traditional end-to-end training frameworks of deep neural networks, which directly learn the mapping from noisy observations to noise-free reconstructions, the proposed DeScoD-ECG learns the conditional distribution with stochasticity. The DeScoD-ECG starts from Gaussian white noise and iteratively reconstructs the signal through a fixed Markov Chain with Gaussian translation conditioned on the previous step reconstructions and the noisy ECG observations. The experimental results demonstrate the potential of the DeScoD-ECG for biomedical signal processing. Compared to the na\"ively training deep neural networks, the diffusion models have better approximations of the true data distribution and higher stability under extreme noise corruptions. The high-quality and high-fidelity ECG reconstruction methodologies could serve the cardiac activity monitoring in various clinical settings, thereby increasing the flexibility of Cardiovascular disease diagnosis. 


\section*{Acknowledgement}
This work was supported by grants from the National Heart, Lung, and Blood Institute (\#R21HL159661), the National Science Foundation (\#2052528 and CAREER \#1943552), and the Department of Energy (\#DE-NA0003946). 
Any opinions, findings, and conclusions or recommendations expressed in this material are those of the authors and do not necessarily reflect the views of the sponsors.



\IEEEpeerreviewmaketitle

\bibliography{my_ref}
\balance
\bibliographystyle{IEEEtran}

\end{document}